# PATHFINDER: Multi-objective discovery in structural and spectral spaces

Kamyar Barakati[1, *], Boris N. Slautin[1], Utkarsh Pratiush[1], Hiroshi Funakubo[2], and Sergei V. Kalinin[1, †]

[1] Department of Materials Science and Engineering, University of Tennessee, Knoxville, TN, USA

[2] Department of Materials Science and Engineering, Institute of Science Tokyo, Yokohama, Japan

**ABSTRACT**

Automated decision-making is becoming key for automated characterization including electron and scanning probe microscopies and nanoindentation. Most machine-learning-driven workflows optimize a single predefined objective and tend to converge prematurely on familiar responses, overlooking rare but scientifically important states. More broadly, the challenge is not only where to measure next, but how to coordinate exploration across structural, spectral, and measurement spaces under finite experimental budgets while balancing target-driven optimization with novelty discovery. Here we introduce PATHFINDER, a framework for autonomous microscopy that combines novelty-driven exploration with optimization, helping the system discover more diverse and useful representations across structural, spectral, and measurement spaces. By combining latent-space representations of local structure, surrogate modeling of functional response, and Pareto-based acquisition, the framework selects measurements that balance novelty discovery in feature and object space and are informative and experimentally actionable. Benchmarked on pre-acquired STEM–EELS data and realized experimentally in scanning probe microscopy of ferroelectric materials, this approach expands the accessible structure–property landscape and avoids collapse onto a single apparent optimum. These results point to a new mode of autonomous microscopy that is not only optimization-driven, but also discovery-oriented, broad in its search, and responsive to human guidance.

[*] kbarakat@vols.utk.edu

[†] Sergei2@utk.edu

I. **INTRODUCTION**

Automated instrumentation is now widely recognized as a prerequisite for self-driving laboratories and materials exploration platforms. In materials characterization, recurring tasks include structural and functional imaging, the discovery of structure–property relationships via coupled imaging and spectroscopy, and multimodal measurements that fuses complementary contrasts or co-orchestrates several characterization tools.[1-5] In principle, these measurements can be integrated across heterogeneous platforms; in practice, a dominant latency bottleneck arises from physically relocating specimens between instruments, with associated alignment and environment-history effects. By contrast, modern electron and scanning probe microscopes already provide rich *in situ* spectroscopic toolkit within a single tool (e.g., EELS/EDS, vibrational/optical probes, and bias- or force-modulated responses), shifting the central challenge from access to efficient use of scarce instrument time withing single imaging platform.[6-8]

In many materials labs, the end-to-end workflow is stable, linear, and slow, with tightly defined transitions between operations.[9,10] A typical example is film growth followed by AFM topography, SQUID magnetometry, and device fabrication and transport measurements. Each step runs under established processes, constrained by cycle times, instrument availability, and qualification requirements.[11,12] In such pipelines, the primary gains come from local optimizations such as throughput on a single tool, recipe tuning, or metrology reliability, without clear need for global, closed-loop orchestration. Because feedback arrives only after hours or days and cannot practically alter upstream steps in real time, ML-driven task planners can only bring marginal value. Often, the primary value of ML and theory workflows is identification of candidate materials, largely decoupled from local process optimization.

By contrast, microscopy platforms including electron and scanning probe are heavily interactive, operate on sub-second to minute feedback loops, and expose rich, high-dimensional control spaces such as probe position, dwell times, or spectroscopy acquisition parameters. In these measurements, the human operator dynamically interacts with the instrument optimizing imaging and spectroscopy parameters, defining regions of interest, switching modalities, and triggering targeted spectra acquisition as hypotheses evolve. Effectively, the human operates constructs the workflow on a fly, enabling the experiment to branch, zoom, or escalate within the same session driving physics discovery. Where feedback is rapid and the action space is broad, ML-enabled orchestration workflows are necessary to effectively use instrument time to acquire structural information of interest, discover physical phenomena and structure–property relations that fixed policy pipelines miss.

On of the central tasks in the context of microscopy is a structure-property discovery via combination of fast low information density imaging and slow high information density microscopy. The examples include STEM-EELS in electron microscopy[13,14], topography imaging and a broad gamut of spectroscopies in SPM[15], nanoindentation assisted by optical or scanning electron microscopy[16], and many others. Here, practical considerations dictate that acquisition of the spectroscopic data over dense spatial grids is both time consuming, associated with possible probe and sample damage, and in some cases impossible due to the destructive nature of measurements. These considerations necessitate development of the active sampling strategies in the image plane, selecting the measurement locations based on expected utility. The same policy-driven framework naturally extends to direct atomic fabrication[17,18] when a reliable causal map exists between probe position/dose and the induced transformation, enabling closed-loop beam placement for structure writing and editing.[19-22]

One practical class of methods targets *a priori* known objects in the field of view. Here, the agent is equipped with fixed or previously learned interest policies that score candidate regions. After an initial survey image, the agent assigns an interest measure to individual features and prioritizes follow-up acquisitions accordingly. Variants of this strategy have been demonstrated in automated SPM, where autonomous workflows have been used to identify suitable surface regions, assess image quality, and selectively revisit targets for higher-resolution imaging or spectroscopy.[23-25] Representative examples include *DeepSPM*, which combined region selection, image-quality assessment, and adaptive probe conditioning into a closed-loop STM workflow[23], as well as autonomous single-molecule SPM imaging, where learned detection and re-identification models were used to locate candidate objects and reacquire them at progressively higher resolution[24]. Related strategies have also been reported for autonomous STM/STS, where defect segmentation and Gaussian-process-guided acquisition were used to prioritize spectroscopic measurements at informative sites rather than exhaustively sampling the full field of view.[25] In electron microscopy, analogous ideas appear in feature-adaptive and dynamically sparse acquisition schemes, in which an initial low-dose image or partial scan is analyzed online to identify regions of structural or chemical interest and to allocate subsequent dwell time or spectroscopic measurements preferentially to those locations.[26]

A second class of approaches focuses on discovering informative targets in the spectral domain. The workflow typically begins with an acquisition of fast structural image and defining property of interest (scalarizer or reward) in from spectral data.[27,28] The ML agent then learns a surrogate model for the structure–property relationship. Using active learning, the controller selects locations to acquire new spectra balancing exploration (minimization of predictive uncertainty) and exploitation (maximizing the reward) goals. Variants of this strategy have been demonstrated in both electron microscopy and scanning probe microscopy.[29-33]

Both fixed policy image-based approaches and reward-based approaches face practical limitations. In the policy-driven, object-finding regime, success depends on the human operator capability to identify features of interest *a priori*, while the full variability of microstructure remains underexplored. While novelty detection algorithms can identify new features, they rely solely on data-driven criteria rather than physical considerations driving the experiment.[34]

In the reward-driven approaches, on-tool implementations frequently converge prematurely, and the learner exploits locally promising regions and fails to traverse broader image spaces. This brittleness is amplified when the reward is only partially specified at run time (e.g., uncertain spectral features-to-property mappings, shifting experimental constraints), so the ML agent optimizes what is easy to score rather than what is scientifically relevant.[35] Moreover, these prior workflows do not extend naturally to multimodality: when imaging is multimodal (multiple contrast channels) and spectroscopy is multimodal, specifying a single, stable policy or scalar reward becomes non-trivial. Hypothesis-learning strategies for multimodal imaging address part of this challenge, but they often assume that all channels are available in parallel, an assumption that breaks down under realistic dose and time budgets or when modalities must be acquired sequentially and re-registered.[36,37]

Human-in-the-loop formulations provide a mechanism for stabilizing autonomous experiment trajectories when the learned surrogate becomes overconfident, is diverted by noisy local optima, or no longer reflects the evolving scientific intent of the measurement. In STEM-EELS, this principle has been formalized in the hAE framework[6], where experiment progression is monitored jointly in real space and latent feature space, enabling intervention when deep-kernel-learning acquisition paths become trapped or cease to accumulate relevant knowledge efficiently. In multimodal and non-differentiable landscapes, the same general problem has been addressed through Strategic Autonomous Non-Smooth Exploration (SANE)[38], which replaces conventional single-optimum acquisition behavior with a probabilistic policy designed to recover multiple high-value regions while incorporating a domain-informed surrogate gate to suppress spurious optima induced by noise. An allied formulation is provided by INS$^2$ANE,[39] which augments strategic exploration with an explicit novelty score so that sampling decisions are not dictated solely by immediate reward maximization, but also by the capacity of a measurement to expand the diversity of observed phenomena. These developments establish that expert intervention, multi-optima search, and novelty-aware acquisition can substantially broaden and stabilize autonomous microscopy workflows; however, they remain embedded within a single adaptive decision layer and therefore do not, by themselves, resolve the more fundamental problem of allocating measurement effort across competing modes of observation or reconciling objectives defined on different data channels.

These limitations point to a more general view of autonomous microscopy, in which the central challenge is not merely where to measure next, but how to coordinate exploration across structural, spectral, and modal degrees of freedom under finite experimental budgets balancing target-driven optimization and novelty discovery in both structural and spectral domains. We therefore recast autonomous experiment design as two coupled tasks: identification of structurally distinct states in the image manifold, and selective assignment of high-value spectroscopic or functional measurements needed to establish their physical significance. This distinction becomes decisive in multimodal experiments, where discovery and interpretation are often supported by different channels and cannot be pursued exhaustively in parallel.

To address this problem, we develop PATHFINDER (Policy for Active, Time-aware, Human-in-the-loop, Fidelity-adaptive Imaging, Navigation, and Discovery with Efficient Rewards), a framework that integrates latent-space characterization of structural novelty with surrogate-based modeling of functional response within a single controller. The algorithm consists of two coupled branches: one quantifying novelty in the structural space from latent representations of local states, and the other modeling response in the spectral domain. Their outputs are combined through multi-objective Bayesian optimization to guide measurement selection under time and dose constraints, while allowing user input through scientific priors and experimental priorities. By linking identification of previously unobserved states to inference of their associated structure–property relations, the framework extends adaptive control beyond a single objective to coordinated selection across structural novelty and functional response. This design helps limit premature convergence and favors measurements that expand coverage of the accessible structure–property landscape while supporting downstream hypothesis generation. The overall architecture is shown schematically in **Fig. 1**.

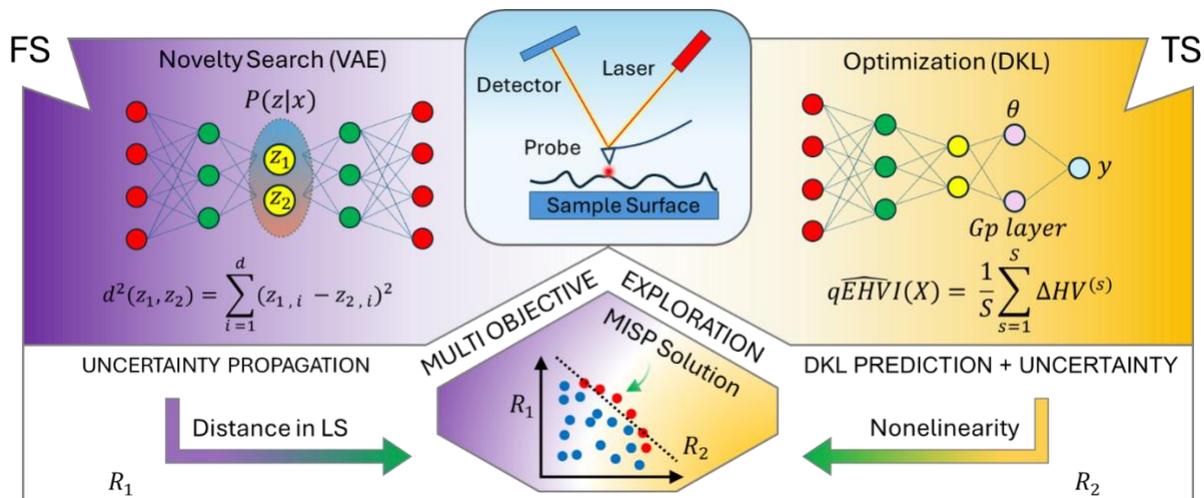

**Figure 1**. The PATHFINDER approach balances the novelty measures in the feature and target space as a multi-objective discovery, providing principled approach for exploring structure-property relationships and human in the loop interventions.

## II. CONCEPTUAL OVERVIEW

The general approach of the workflow is chosen to fully represent and automate the human-based decision making when operating the microscope, specifically selecting the regions for the detailed spectroscopic studies. Generally, human decision making in this case can be driven either by the selection of certain objects of interest in the structural image, or by the search for specific spectroscopic signatures.

For the structural image, human operators can select objects of interest based either on the prior knowledge, statistically significant features, or some form of subjective novelty. The structural image is generally available before selection of the objects of interest for spectroscopic studies is performed. It is important to note that the human definition of objects of interest can evolve during the experiment as new data becomes available during experiment, or the operator queries external information sources or gains advice from colleagues or online sources. Correspondingly, exploring the objects in the image plane requires some definition of "interest" that can be introduced as a reward function that can be updated during the experiment, with the relative sparsity or structural complexity being the simplest possible definitions.

The second set of reward functions are associated with the spectroscopic measurements. Unlike structural descriptors, the spectroscopic measurements are not available in the beginning of the experiment, and locations are chosen sequentially based on the fully available structural data and partially available spectral data. In this case, the exploratory strategies require defining both the reward function specifying the interest to spectral data (scalarizers) and learning the correlative relationship between structural descriptors and the scalarizers. Over the last several years, these workflows have been realized via versions of Deep Kernel Learning (DKL) within the Bayesian Optimization cycles, but other surrogate models can be created. Like structural descriptors, the scalarizers are known only probabilistically and can be dynamically updated during the experiment based on the data obtained up to specific moment.[27,31,40-42]

Once the reward functions are defined, the exploration strategy must balance exploration and exploitation. While for human led experiments these are often defined only implicitly, in the ML workflows the introduction of acquisition functions allows to define them rigorously. Importantly, the exploration-exploitation balance can be dynamically updated during the experiment giving rise to the human in the loop experimental frameworks.

Over the last 5 years, multiple frameworks for fixed policy, fixed reward in the structural and spectral domain, and human in the loop automate experiment has been proposed as summarized in the introduction. The algorithm offers the full generalization of this algorithm by dynamically balancing discovery in the structural and spectral domains and allow the hAE intervention in terms of reward, policies, and exploration and exploitation balance.

### II.1. STRUCTURAL REWARD $R_1$

To develop intuition behind algorithm, we first consider the case of fully available structural and spectral data, illustrated here for the STEM-EELS data set. Here, both the structural patches and the spectra are available at all spatial locations. Such pre-acquired data sets are ideal both to benchmark the algorithm as instrument emulators, and to generate ground truth answers. It is important to note though that experimentally the limitations on possible data volumes usually limit their applicability to only short experimental campaigns.[35, 43]

In this study, we define the structural reward, $R_1$, as novelty in feature space. Specifically, we define the novelty as the distance of measured point from previously measured points in the latent space of the variational autoencoder (VAE) representation of the image patches. The use of the VAE for building compact representations of complex microstructures has been explored.[44-48] In this setting, latent-space distance provides a natural measure of structural novelty, although other definitions could also be adopted, for example similarity to a selected object of interest when such external information is available.

### II.2. FUNCTIONAL REWARD $R_2$

We further define the spectral or functional reward, $R_2$, as the target response modeled by the surrogate. Depending on the experimental modality, this response may correspond to EDS or EELS signals in STEM, nonlinearity or hysteresis-loop observables in AFM, or other measurement-specific quantities of interest. In this way, $R_2$ captures the functional objective, whereas $R_1$ promotes exploration of structurally distinct states. These two rewards are combined through multi-objective Bayesian optimization to guide measurement selection.

### II.3. WORKFLOW

The rationale of the workflow is that optimal experimental outcomes in heterogeneous systems need not arise from the most prevalent structural class. Accordingly, the full workflow leverages contextual information to define a prior over structural heterogeneity, rather than using candidate locations solely as coordinates for sampling. By embedding local candidates into a latent space and defining $R_1$ through latent-space isolation, the method preferentially emphasizes rare or weakly represented motifs, thereby introducing structural informativeness before costly measurement.

Experimental feedback enters through a second objective, $R_2$, which is observed only at sampled locations. A surrogate model is updated sequentially on measured $R_2$ values to infer the expected response and its uncertainty over the unsampled set. The central design choice is to retain $R_1$ and $R_2$ as separate objectives, since structural distinctiveness and experimental utility are not necessarily aligned.

Candidate selection is performed by maximizing expected hypervolume improvement in the joint $(R_1, R_2)$ space. This favors points that expand the Pareto front between structural novelty and experimental performance, rather than maximizing either objective in isolation. As a result, the workflow the workflow described in **Algorithm 1** avoids premature convergence to a single apparent optimum and instead promotes efficient discovery of diverse, high-value structure–property relationships under limited experimental budget.

---

**Algorithm 1**: Contextual reward-driven microscopy with VAE structural prior

Input:
  I. Map amplitude Channel : $I : \Omega \rightarrow \mathbb{R}$
  II. Patch Size: $w$
  III. Stride: $s$
  IV. Seed budget: $n_0$
  V. Total budget: $T$

Steps:
1. Construct the candidate patch set $\mathcal{P} = \{P_i\}_{i=1}^{N}$ by sliding-window sampling of $I$, with $P_i = \mathcal{W}(I; c_i)$, where $c_i$ is the center of patch $i$.
2. **Train a VAE encoder $q_\phi(z \mid P)$ on $\{P_i\}_{i=1}^{N}$.**
3. For each candidate $i \in [N]$, encode patch $P_i$ and compute the latent embedding
   $q_\phi(z_i \mid P_i) = \mathcal{N}(\mu_i, \Sigma_i)$,
   $z_i = \mu_i \in \mathbb{R}^m$.
4. Compute the structural reward map $R_1$ over all candidates. For example,
   $\bar{z} = \frac{1}{N} \sum_{i=1}^{N} z_i$,
   $R_1(i) = \|z_i - \bar{z}\|_2$,
   or, alternatively,
   $R_1(i) = \frac{1}{k} \sum_{j \in \mathcal{N}_k(i)} \|z_i - z_j\|_2$
   where $\mathcal{N}_k(i)$ denotes the k nearest neighbors of $z_i$ in latent space
5. Initialize the measured set $S_0$ by selecting $n_0$ seed locations
6. For each $i \in S_0$, perform experiment $\mathcal{M}(i)$ and obtain the measured reward
   $R_2(i) = \mathcal{M}(i)$
7. Form the initial dataset
   $\mathcal{D}_0 = \{(P_i, R_1(i), R_2(i)) : i \in S_0\}$
8. For $t = 1, \ldots, T$ do
   a. Fit a surrogate model for the experimentally observed reward $R_2$ using $\mathcal{D}_{\{t-1\}}$, and compute, for all $i \in [N]$,
      $\widehat{R}_2^{(t)(i)}, \widehat{\sigma}_2^{(t)(i)}$.
   b. For each candidate $i \notin S_{\{t-1\}}$, form the predictive objective statistics
      $\mu_i^{(t)} = \begin{bmatrix} R_1(i) \\ \widehat{R}_2^{(t)(i)} \end{bmatrix}, \Sigma_i^{(t)} = \begin{bmatrix} \widehat{\sigma}_1(i)^2 & 0 \\ 0 & \widehat{\sigma}_2^{(t)}(i)^2 \end{bmatrix}.$
      Equivalently,
      $Y_i^{(t)} \sim \mathcal{N}(\mu_i^{(t)}, \Sigma_i^{(t)})$.
   c. Let
      $\mathcal{Y}_{\{t-1\}} = \{(R_1(j), R_2(j)) : j \in S_{\{t-1\}}\}$
   d. For each candidate $i \notin S_{\{t-1\}}$, define the hypervolume improvement
      $AHV_t(i : y) = HV(\mathcal{Y}_{\{t-1\}} \cup \{y\}; r) - HV(\mathcal{Y}_{\{t-1\}}); r$,

and evaluate the acquisition score
$$\alpha_{t(i)} = E\left[AHV_t\left(i; Y_i^{(t)}\right)\right]$$
  e. Select the next location $i_t = argmax_{i \notin S_{\{t-1\}}} \alpha_{t(i)}$
  f. Perform experiment at $i_t$ and measure $R_2(i_t) = \mathcal{M}(i_t)$
  g. Augment the measured set and dataset:
$$S_t = S_{\{t-1\}} \cup \{i_t\}$$
$$D_t = D_{\{t-1\}} \cup \{(P_{i_t}, R_1(i_t), R_2(i_t))\}$$
  h. end for
9. Return the visited trajectory $S_T$, the measured rewards $\{R_1(i), R_2(i)\}_{i \in S_T}$, the surrogate reward maps, and the Pareto set in $(R_1, R_2)$-space.

## III. MATERIALS SYSTEMS AND EXPERIMENTAL SETTINGS

To evaluate the framework in both fully observed and sequential autonomous settings, we consider two complementary experimental systems. The first is a pre-acquired STEM–EELS dataset of fluorine- and tin-co-doped indium oxide nanoparticles with near-infrared plasmon resonances.[49-52] This system provides a structurally and spectrally heterogeneous landscape in which both the HAADF image and the corresponding spectral response are available throughout the field of view, making it well suited as a benchmark for analyzing joint structural–spectral discovery under controlled conditions. The second is an on-the-fly scanning probe experiment on epitaxial $PbTiO_3$ thin films grown on (100) $KTaO_3$ single-crystal substrates. In this heterostructure, epitaxial tensile strain stabilizes the in-plane ferroelectric state and produces pronounced spatial variations in local electromechanical response, providing a rich landscape for active interrogation. Unlike the pre-acquired benchmark, this setting begins from a single structural scan, after which the functional response is obtained sequentially during experiment progression. Together, these two systems allow us to examine the framework both in a fully observed benchmark setting and in a live autonomous experiment.

### III.1. PRE-ACQUIRED BENCHMARK: STEM–EELS NANOPARTICLE DATA

The pre-acquired STEM–EELS benchmark consists of a nanoparticle dataset in which both the structural image and the corresponding spectral response are available across the full field of view. A representative HAADF image of the ensemble is shown in **Fig. 2a**.

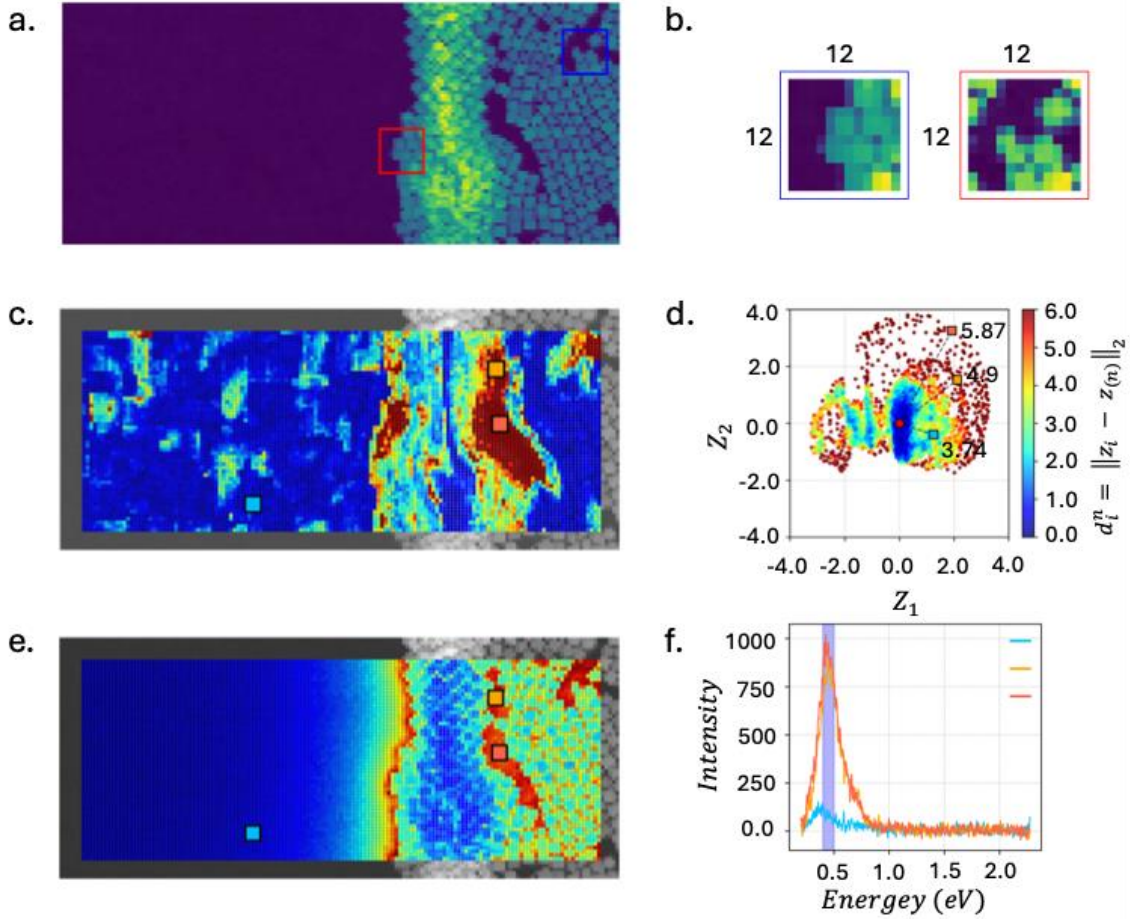

*Figure 2: Structural novelty and spectral reward in real and latent spaces. a) Representative HAADF image of the nanoparticle sample. Colored boxes indicate two examples 12 × 12 pixel image patches used for structural analysis. b) Enlarged views of the corresponding structural patches. c) Spatial map of the structural novelty reward $R_1$, computed from the VAE latent representation of local patches using the neighbourhood-based novelty metric. Colored squares mark representative locations discussed in d–f. d) Distribution of the same patches in the two-dimensional latent space ($Z_1$, $Z_2$), colored by the corresponding structural novelty score. The marked points correspond to the locations highlighted in c. e) Spatial map of the spectral reward $R_2$, defined from the scalarized EELS response and overlaid on the structural image. f) EELS spectra extracted from the locations marked in c and e; the shaded region denotes the energy window used to define $R_2$. Together, these panels illustrate the structural organization of the sample in both real and latent spaces, and the complementary variation of the spectral reward across the same field of view.*

To parameterize local structure, the HAADF field of view shown in **Fig. 3a** was decomposed into overlapping 12 × 12 pixel patches, with two representative examples shown in **Fig. 3b**. Each patch was encoded by a pretrained variational autoencoder (VAE) into a latent vector $z_i \in \mathbb{R}^m$, such that Euclidean distances in latent space quantify structural similarity. This latent representation defines the structural reward $R_1$. In the main-text analysis, $R_1$ was taken as a local novelty measure, computed from the separation of a patch from a reference point identified within its latent-space neighborhood. The resulting spatial map of structural novelty is shown in Fig. 3e, and its distribution in the two-dimensional latent embedding is shown in **Fig. 3f**. For comparison, **SI Fig. S1** presents both reward definitions considered here. In **SI Fig. S1a,b,** novelty is defined relative to the global latent mean, $\bar{z} = \frac{1}{N}\sum_{i=1}^{N} z_i$, such that $d_i^{mean} = \parallel z_i - \bar{z} \parallel_2$, whereas **SI Fig. S1c, d** show the neighborhood-based definition used in the main text, $d_i^{(n)} = \parallel z_i - z_{(n)} \parallel_2$. These two constructions provide complementary notions of structural novelty, corresponding to deviation from the dominant latent manifold and dissimilarity at the neighborhood scale, respectively.

The second objective, $R_2$, was defined directly from the EELS response. For a spectrum $s_i(E)$ acquired at spatial position $i$, we define $R_2(i) = \int_{E_{min}}^{E_{max}} s_i(E)\, dE$, where $[E_{min}, E_{max}]$ denotes the energy window associated with the spectral feature of interest. In discrete form, this is equivalently written as the mean or summed intensity over the selected energy interval. The spatial distribution of this spectral reward is shown in **Fig. 3e**, overlaid on the structural image, and three representative spectra extracted from the marked positions are displayed in **Fig. 3f**. The shaded interval in **Fig. 3f** indicates the energy range used for scalarization.

We formulated the experiment as a discrete multi-objective Bayesian optimization problem over all $(7755, 12, 12)$ HAADF patches extracted with unit stride from the field of view in **Fig. 3a**. Each candidate location was assigned two rewards: a structural novelty score $R_1$, derived from the pretrained 2D VAE latent embedding of the local patch, and a spectral score $R_2$, defined from the scalarized EELS intensity in the 0.4–0.5 eV range shown in **Fig. 3d**. Unless noted otherwise, the structural objective was taken to be the neighborhood-based novelty metric shown in **Fig. 3e,f** and **SI Fig. S1c, d**; the alternative global-mean metric is reported in **SI Fig. S1a,b**. Separate deep-kernel Gaussian-process surrogates were trained for the two objectives using convolutional patch embeddings, and the next query location was selected by maximizing a hypervolume-improvement acquisition function over the remaining candidates. The experiment was initialized from a random seed set of 10 locations, after which the measured set, Pareto front, and surrogate models were updated iteratively under a fixed experimental budget. The same workflow was also performed with $R_1$ defined as the distance from the global latent mean in VAE space; the corresponding results are shown in the **SI Fig. S2**.

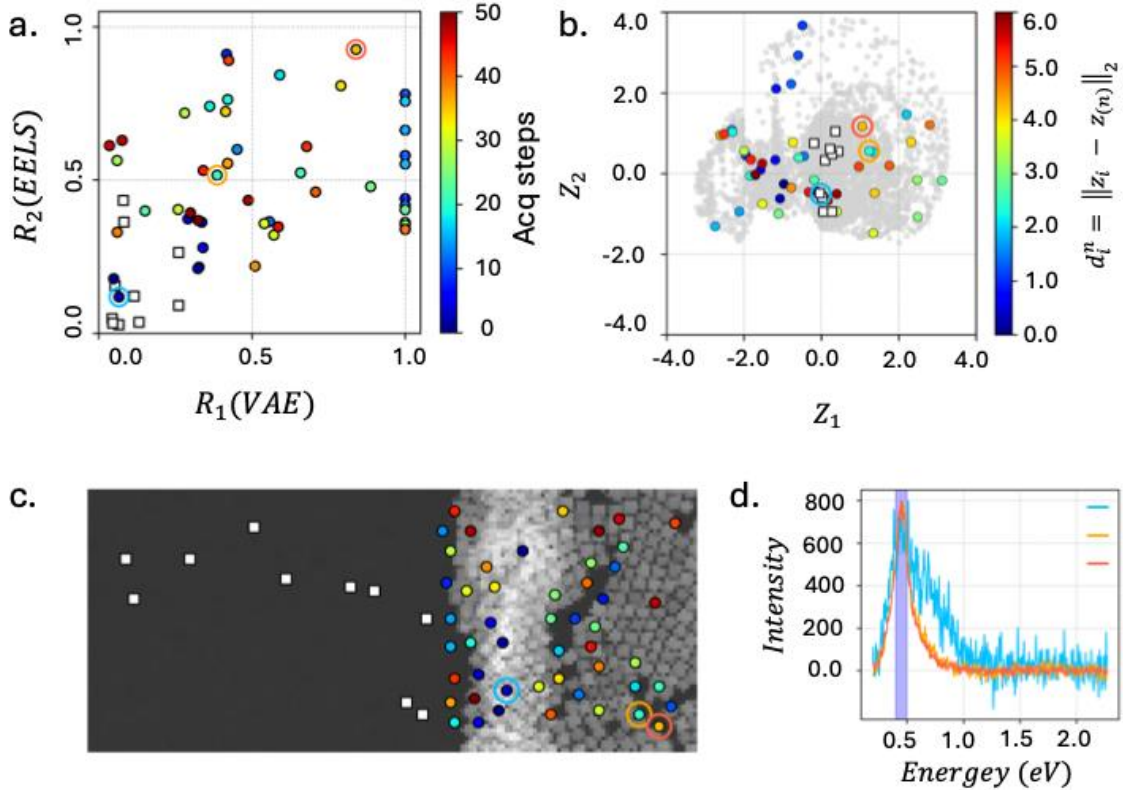

*Figure 3*: Multi-objective autonomous experiment in structural and spectral spaces. a) Pareto-optimal solutions in the two-objective reward space defined by the structural reward $R_1$ derived from the VAE latent representation and the spectral reward $R_2$ derived from the EELS response. Marker color denotes acquisition step. b) Corresponding trajectory of the selected measurements in the VAE latent space. c) Spatial distribution of the measured locations overlaid on the structural image. d) Representative EELS spectra acquired at the circled locations in c, with the shaded band indicating the energy window used to define $R_2$.

The optimization trajectory reveals a search process that remains exploratory while progressively concentrating on high-value measurements. In the joint reward space (**Fig. 3a**), successive acquisitions populate a broad set of Pareto-optimal trade-offs between structural novelty, $R_1$, and spectral reward, $R_2$, rather than collapsing rapidly onto a single corner of the objective space. This behavior is mirrored in the VAE latent embedding (**Fig. 3b**), where queried points continue to sample distinct structural regions of the manifold, indicating that convergence in reward space is achieved without loss of structural coverage. In real space (**Fig. 3c**), the selected locations are distributed across the field of view, with repeated attention to the interfacial region while still retaining measurements in more weakly varying areas, consistent with a balance between refinement and continued exploration. The spectra extracted from representative queried points (**Fig. 3d**) confirm that these locations correspond to distinct EELS responses within the scalarized energy window, linking the spatial and latent-space trajectories to measurable functional contrast. Taken together, this workflow indicates that the search does not proceed by simple exploitation of a single apparent optimum but instead develops along a structured frontier that preserves diversity in local structure while enriching for regions with strong spectral response.

III.2. ON-THE-FLY EXPERIMENT: FERROELECTRIC SPM

The on-the-fly ferroelectric experiment was initialized from a single structural scan of the $PbTiO_3$ film surface. The image was normalized and tessellated into overlapping $20 \times 20$ patches with a stride of 5 pixels, and the center of each patch was converted from image coordinates into real-space probe coordinates after correction for scan rotation. Each candidate measurement was thus represented by both a local structural patch and an experimentally accessible tip position. A two-dimensional invariant variational autoencoder was trained on the full patch library to embed every candidate into a latent structural manifold $z_i$. A visual summary of this experimental setup, including the measurement map, latent-space embedding, and structural reward representations, is provided in SI **Figure S3**.

In this case, novelty was referenced to the entire latent manifold available at initialization, for example through a neighborhood-isolation statistic evaluated over all embedded points. In the adaptive formulation, $R_1$ was defined only with respect to the set of previously measured locations. If $\mathcal{M}_t$ denotes the set of measurements available after iteration $t$, then the novelty of a candidate $z_i$ was evaluated relative to $\mathcal{M}_t$, for example as $R_1(i,t) = \max_{j \in \mathcal{M}_t} \| z_i - z_j \|_2$, or by an analogous neighbourhood statistic restricted to the measured set. Thus, immediately after seeding, the structural reward was determined only by the initial seed points; after each subsequent acquisition, the reference set expanded and the novelty landscape was updated accordingly. The static formulation therefore measures rarity with respect to the global structural manifold, whereas the adaptive formulation promotes progressive exploration away from the trajectory already sampled by the controller.

The functional reward was obtained from local force spectroscopy and defined as the absolute value of the recovered nonlinearity coefficient, $R_2(i) = | \eta_i |$. Starting from 10 random seed measurements, the experiment proceeded for 50 autonomous acquisitions. At each iteration, a deep-kernel Gaussian-process surrogate was fit to the nonlinear-response objective, expected hypervolume improvement was evaluated over the remaining candidates, and the next coordinate was selected to maximize Pareto improvement in the joint $(R_1, R_2)$ space. In this way, the workflow operated as a multi-objective, budget-aware controller that balanced structural exploration against functional discovery.

The behavior of the active experiment is summarized in **Fig. 4**. In real space **Fig. 4a**, the acquisition trajectory is clearly non-uniform, concentrating instead on specific domain configurations, junctions and interfacial regions. This selective sampling reflects the central design of the method: measurements are directed not by geometric coverage alone, but by their expected contribution to the joint structural–functional objective. The corresponding trajectory in the learned latent manifold **Fig. 4b** spans multiple separated regions of the VAE space, indicating that the controller does not collapse rapidly onto a single structural class, but continues to interrogate diverse local environments throughout the run. The origin of this behavior is evident in the structural novelty landscape **Fig. 4c**. High values of $R_1$ occupy sparsely populated regions of latent space, corresponding to patches that are structurally uncommon with respect to the dominant manifold. When projected back into real space **Fig. 4d**, these states localize preferentially near domain boundaries and heterogeneous contrast features, showing that the latent representation captures physically relevant microstructural variation rather than incidental fluctuations in image intensity. In this sense, the structural branch of workflow acts to preserve access to rare or weakly represented local states during experiment progression. The functional branch exhibits a distinct organization. The surrogate prediction for the nonlinear-response objective $R_2$, **Fig. 4e** reveals a spatial pattern that only partially overlaps with the structural novelty map, demonstrating that structurally unusual regions are not, in general, identical to those with the largest functional response. This partial decoupling is precisely what necessitates a multi-objective controller: neither structure nor function alone is sufficient to define the most informative next measurement. The acquisition landscape **Fig. 4f** shows how these two objectives are reconciled. The high-acquisition regions are sparse and sharply localized, indicating that the next query is not chosen simply from maxima of $R_1$ or $R_2$ separately. Instead, the workflow selects points expected to yield the largest improvement in Pareto hypervolume, that is, points most likely to reveal new and informative structure–function trade-offs. The resulting trajectory therefore reflects adaptive discovery: each measurement updates the surrogate, reshapes the acquisition landscape and redirects the experiment toward regions that remain both structurally informative and functionally promising. The same workflow is shown for the static measurement protocol in the SI **Figure S4**.

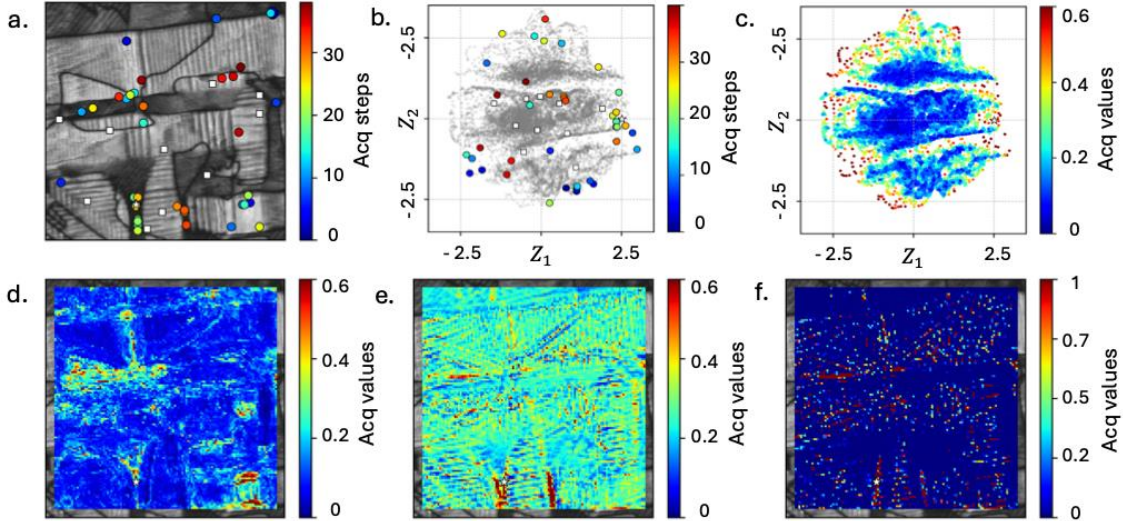

*Figure 4: Active realization of the workflow in structural and functional spaces. a) Real-space acquisition trajectory overlaid on the structural image; coloured markers denote autonomous measurements and white squares indicate the initial seed points. b, Corresponding trajectory in the VAE latent space. c) Structural novelty landscape in latent space, defined by the VAE-based reward $R_1$. d) Real-space projection of the structural reward $R_1$. e) Surrogate prediction of the functional reward $R_2$ associated with the nonlinear spectroscopic response. f) Acquisition landscape given by expected hypervolume improvement, highlighting locations predicted to yield the largest gain in Pareto-optimal structure–function trade-offs.*

The stepwise evolution is summarized in **Fig. 5**. In the joint reward space **Fig. 5a**, the initial seed measurements (white squares) sample only a limited portion of the accessible $(R_1, R_2)$ landscape. Subsequent autonomous acquisitions (colored points) progressively expand this coverage towards the Pareto-relevant region, indicating that the controller does not optimize novelty or functionality in isolation, but refines the trade-off between them. Early steps broaden structural coverage, whereas later steps increasingly populate points that combine high VAE-space novelty with enhanced nonlinear response. The corresponding spectra (**Fig. 5b**) confirm that points selected at different stages of the run exhibit distinct nonlinear behavior, showing that the trajectory continues to diversify the measured set rather than repeatedly sampling equivalent states. The corresponding analysis for the static experiment is provided in **SI Figure S5**, which shows the same evolution of measured points and the associated spectroscopic responses.

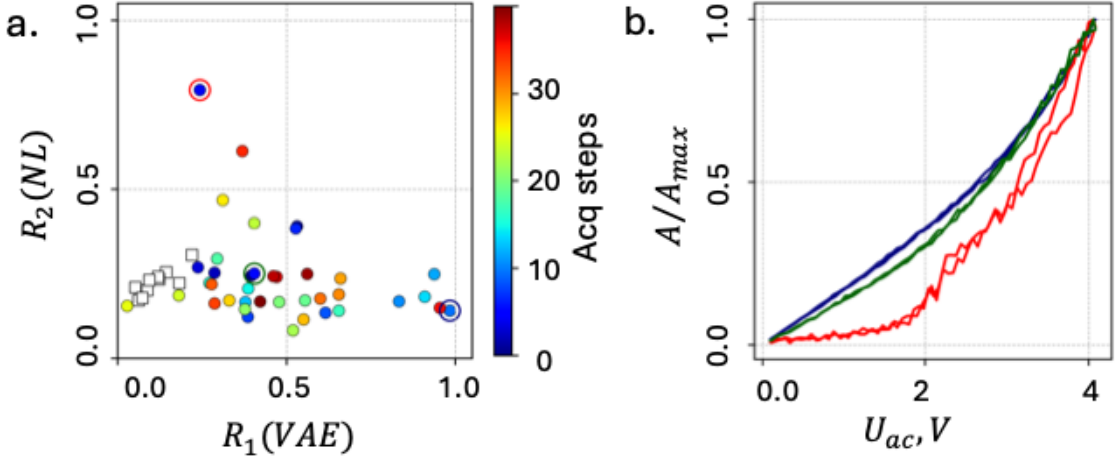

*Figure 5*: Stepwise evolution of the active experiment. a) Measured points in the joint reward space $(R_1, R_2)$, where grey squares denote the initial seeds and coloured points indicate subsequent acquisitions; circled markers highlight representative measurements. b) Corresponding nonlinear responses for the circled points in a, showing distinct spectroscopic behaviors sampled during the experiment.

## IV. DISCUSSION

The contrast between the static formulation, where novelty is computed relative to the full set of pre-acquired points, and the adaptive formulation, where novelty is computed only from the set of points measured up to the current step, becomes most evident in the cumulative latent-space coverage shown in **Fig. 6a**. The adaptive formulation achieves a systematically larger explored fraction of the structural manifold over the same acquisition budget, indicating that redefining novelty with respect to the measurement history materially changes the trajectory of the experiment. In the static case, structural rarity is fixed at initialization; once a high-value region has been identified, nearby states can remain attractive because their novelty is referenced to the global latent distribution rather than to the path already taken. In the adaptive case, by contrast, each new acquisition changes the reference set itself, so that recently visited neighborhoods lose relative priority. The controller is therefore driven away from already sampled basins and toward structurally unvisited regions, producing broader manifold coverage. This distinction is also reflected in the stepwise evolution of the functional reward $\alpha_{NL}$ in **Fig. 6b, c.** Neither trajectory is monotonic. Instead, both exhibit short intervals of local increase followed by partial decline. This pattern is consistent with an uncertainty-aware multi-objective search: once a region of elevated nonlinear response has been located, several successive measurements may refine that local optimum, giving rise to a transient upward trend. However, once the marginal gain from that neighborhood decreases, the acquisition rule redirects the search toward alternative regions that may offer a better overall trade-off between structure and function. The resulting fluctuations should therefore not be interpreted as instability, but as the expected signature of a controller that alternates between local refinement and global exploration. The same logic is apparent in the latent-space displacement between consecutive measurements, $z_{t+1} - z_t$, shown in **Fig. 6d, e.** In the static mode **Fig. 6d**, large jumps are interspersed with intervals of shorter motion, indicating that the experiment can remain within favorable structural neighborhoods before relocating. In the adaptive mode **Fig. 6e**, large displacements occur more persistently, consistent with the fact that each acquisition reduces the novelty of its local surroundings and thereby suppresses redundant resampling. The adaptive trajectory is therefore more path-dependent: measurement does not simply add information but actively reshapes the objective landscape that governs the next decision. Taken together, **Fig. 6a–e** show that the higher coverage of the adaptive formulation does not arise from random wandering, but from a principled redefinition of novelty during experiment progression. Smooth, saturating reward traces would be expected from a purely exploitative optimizer that remains locked to one dominant response. The jagged but locally increasing trajectories observed here instead indicate repeated entry into promising regions, partial local refinement, and subsequent departure once those regions cease to expand the attainable trade-off surface. In this sense, the adaptive formulation is not merely more exploratory; it is more effective at converting finite measurement budget into a broader sampling of accessible structure–function states.

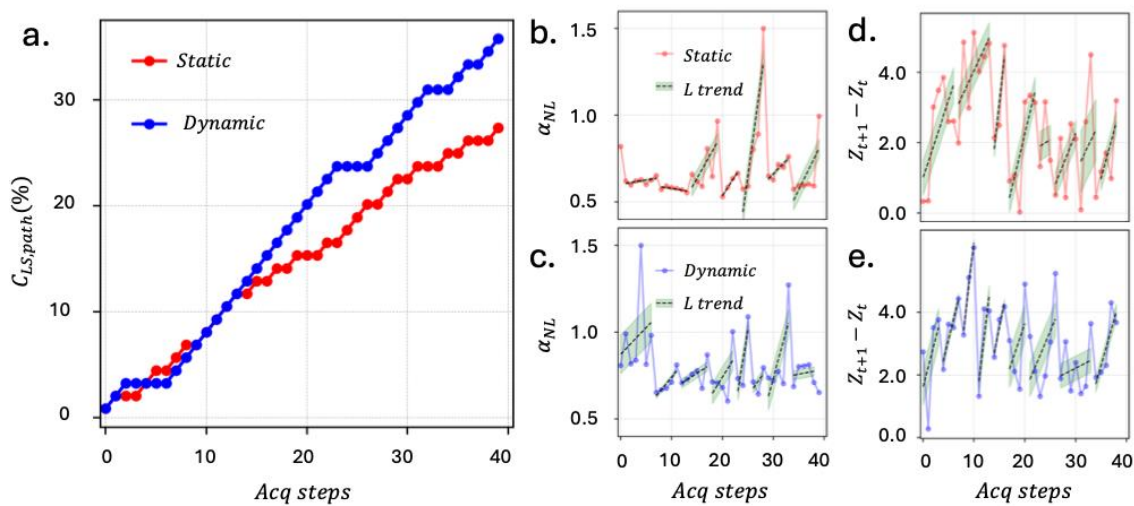

*Figure 6*: Coverage and stepwise reward evolution in static and adaptive novelty formulations. a) Cumulative latent-space coverage, $C_{LS,path}$, as a function of acquisition step for the static (red) and dynamic (blue) modes. b, c) Evolution of the nonlinear-response objective, $\alpha_{NL}$, during the static and dynamic runs, respectively; dashed lines indicate local

*trends. d, e) Consecutive latent-space displacement, $z_{t+1} - z_t$, for the static and dynamic runs, respectively, showing the balance between local refinement and redirection to structurally distinct regions.*

**SUMMARY AND OUTLOOK**

By coupling structural novelty with functional optimization, this work outlines a broader experimental logic for autonomous microscopy: one that does not simply pursue the strongest immediate response but instead allocates measurement effort across diverse local states to uncover informative structure–property relationships under finite budget. In both the STEM–EELS benchmark and the active ferroelectric microscopy experiments, the framework expands exploration of the accessible landscape while retaining the ability to refine regions of high functional interest, thereby avoiding the premature collapse that often accompanies single-objective acquisition. The distinction between static and adaptive novelty further shows that redefining structural reward during experiment progression can materially improve coverage of the latent manifold and redirect the search towards previously unvisited regimes. A natural next step is to extend this formulation to genuinely multimodal and multi-fidelity settings, where imaging, spectroscopy and external priors are integrated within a common decision loop and where reward, uncertainty and cost are updated jointly in real time. In that regime, autonomous microscopy would move beyond local optimization towards a more general, hypothesis-driven mode of experimentation that is discovery-oriented, physically grounded and responsive to evolving scientific intent.

## DATA AVAILABILITY

Additional workflow details and visual summaries are provided in the Supplementary Information, including representative realizations of the experimental setup and acquisition maps

## CODE AVAILABILITY

The code used to generate and analyze the datasets during the current study is available at GitHub.

## ACKNOWLEDGEMENTS

This work (workflow development and conceptualization) was supported (K.B., B.S., and S.V.K.) by the U.S. Department of Energy, Office of Science, Office of Basic Energy Sciences, through the Energy Frontier Research Centers program: CSSAS—The Center for the Science of Synthesis Across Scales (Award No. DE-SC0019288), which supported prototyping on EELS, and the Center for 3D Ferroelectric Microelectronics Manufacturing (3DFeM2) (Award No. DE-SC0021118), which supported implementation in the automated experiment. The work was partially supported (U.P.) by the AI Tennessee Initiative at the University of Tennessee, Knoxville. Film growth (H.F.) was supported by MEXT Program: Data Creation and Utilization Type Material Research and Development Project (No. JPMXP1122683430) and MEXT Initiative to Establish Next-generation Novel Integrated Circuits Centers (X-NICS) (JPJ011438), and the Japan Science and Technology Agency (JST) as part of Adopting Sustainable Partnerships for Innovative Research Ecosystem (ASPIRE), Grant Number JPMJAP2312.

## COMPETING INTERESTS

The authors declare no competing interests.